\documentclass[aps,prl,twocolumn,superscriptaddress,groupedaddress]{revtex4}  
\usepackage{graphicx}  
\usepackage{bm}        
\usepackage{amsmath}
\usepackage{amsfonts}
\usepackage{amssymb}
\usepackage{subfigure}

\renewcommand{\vec}[1]{\boldsymbol{#1}}
\newcommand{\tens}[1]{\boldsymbol{#1}}

\newcommand{\fig}[2]{{\bf Fig.~\ref{#1}#2}}

\newcommand{\mov}[1]{{\bf Movie~{#1}}}

\begin{document}

\widetext
\leftline{Accepted to \textit{Science Advances} on 14/06/2016}

\title{Active micromachines:\\{M}icrofluidics powered by mesoscale turbulence}

\author{Sumesh P. Thampi}
\affiliation{Department of Chemical Engineering, Indian Institute of Technology Madras, Chennai 600036, India.}
\affiliation{The Rudolf Peierls Centre for Theoretical Physics, 1 Keble Road, Oxford, OX1 3NP, United Kingdom.}
\author{Amin Doostmohammadi}
\affiliation{The Rudolf Peierls Centre for Theoretical Physics, 1 Keble Road, Oxford, OX1 3NP, United Kingdom.}
\author{Tyler N. Shendruk}
\affiliation{The Rudolf Peierls Centre for Theoretical Physics, 1 Keble Road, Oxford, OX1 3NP, United Kingdom.}
\author{Ramin Golestanian}
\affiliation{The Rudolf Peierls Centre for Theoretical Physics, 1 Keble Road, Oxford, OX1 3NP, United Kingdom.}
\author{Julia M. Yeomans}
\email[Corresponding author: ]{julia.yeomans@physics.ox.ac.uk}
\affiliation{The Rudolf Peierls Centre for Theoretical Physics, 1 Keble Road, Oxford, OX1 3NP, United Kingdom.}

\date{\today}

\begin{abstract}
Dense active matter, from bacterial suspensions and microtubule bundles driven by motor proteins to cellular monolayers and synthetic Janus particles, is characterised by mesoscale turbulence, the emergence of chaotic flow structures. By immersing an ordered array of symmetric rotors in an active fluid, we introduce a microfluidic system that exploits spontaneous symmetry breaking in mesoscale turbulence to generate work. The lattice of rotors self-organises into a spin-state where neighbouring discs continuously rotate in permanent alternating directions due to combined hydrodynamic and elastic effects. Our virtual prototype demonstrates a new research direction for the design of micromachines powered by the nematohydrodynamic properties of active turbulence.
\end{abstract}

\maketitle

\section{Introduction}

A hallmark of active fluids, such as bacterial suspensions and filament-motor protein mixtures, is the emergence of large-scale chaotic, turbulent-like structure known as mesoscale (or active) turbulence~\cite{Julia2012,Dogic2012,Silberzan2010,Vedula2012}. The intrinsic disorder of these flow fields acts as a barrier to the extraction of useful power from active suspensions.  Pre-designed microfluidic systems have been built to direct cell motion~\cite{Weibel2005,Austin2007,Shear2009,Mahmud2009}, pump fluid~\cite{Berg2004} or drive microrotors~\cite{Hiratsuka2006,Vogel2008}.  These impose symmetry breaking through geometrical constraints or external biases. In particular, it is possible to obtain a persistent rotation for specially designed microscopic gears submersed in bacterial baths~\cite{Leonardo2009,Leonardo2010,Sokolov2010}. However, it has been argued that an asymmetric gear shape is required to achieve a persistent rotation and bacterial “traps” in the form of symmetric teeth are essential for generating directed motion of gears~\cite{Leonardo2010}. Here, we argue asymmetry of the gears is not a prerequisite and propose a means of exploiting spontaneous symmetry breaking of mesoscale turbulence to turn many symmetric rotors.

\begin{figure*}
	\centering
	\subfigure[]{\includegraphics[width=0.95\textwidth]{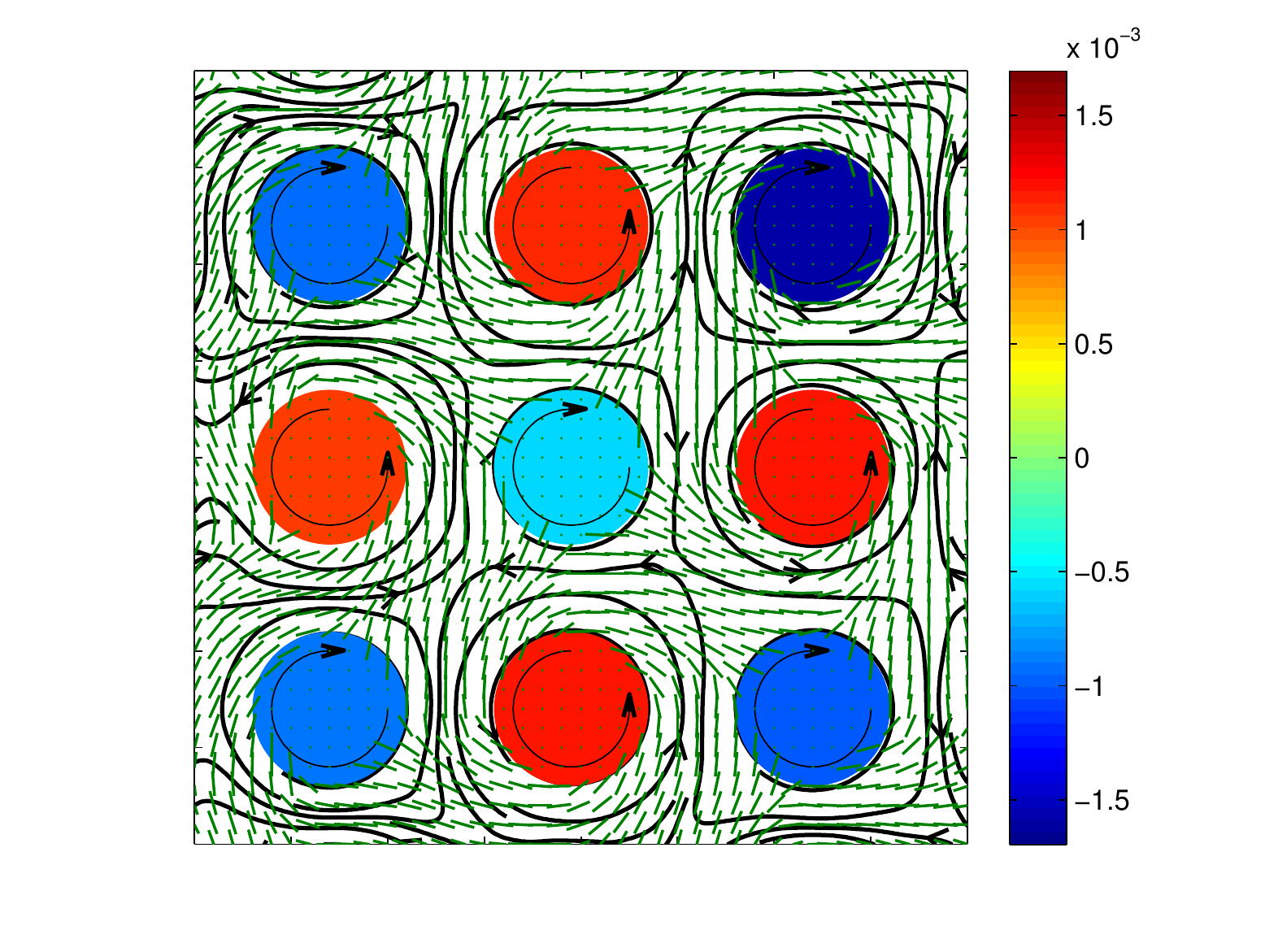}} \label{fig1a}
	\\
	\subfigure[]{\includegraphics[width=0.45\textwidth]{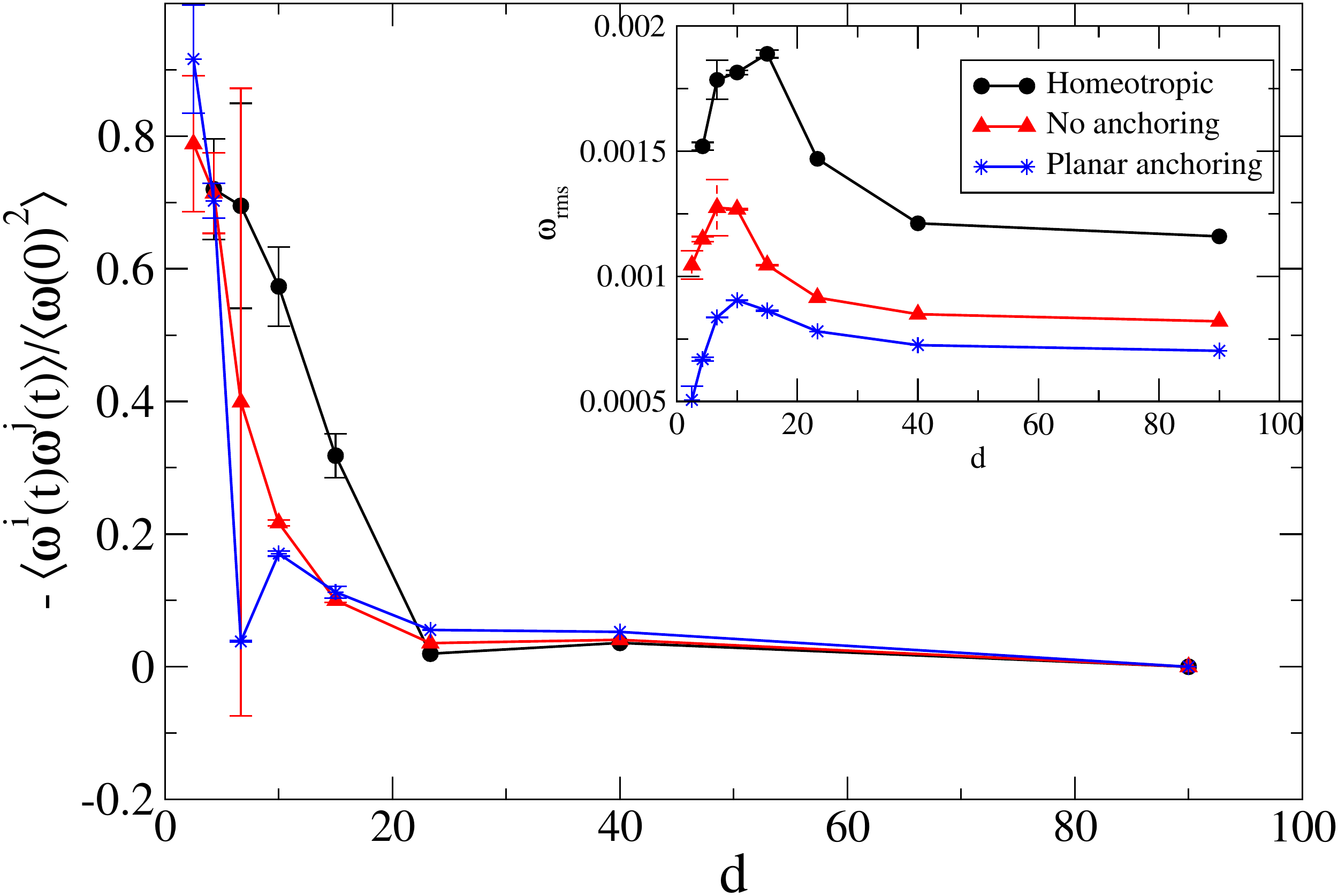}} \label{fig1b}
	~
	\subfigure[]{\includegraphics[width=0.45\textwidth]{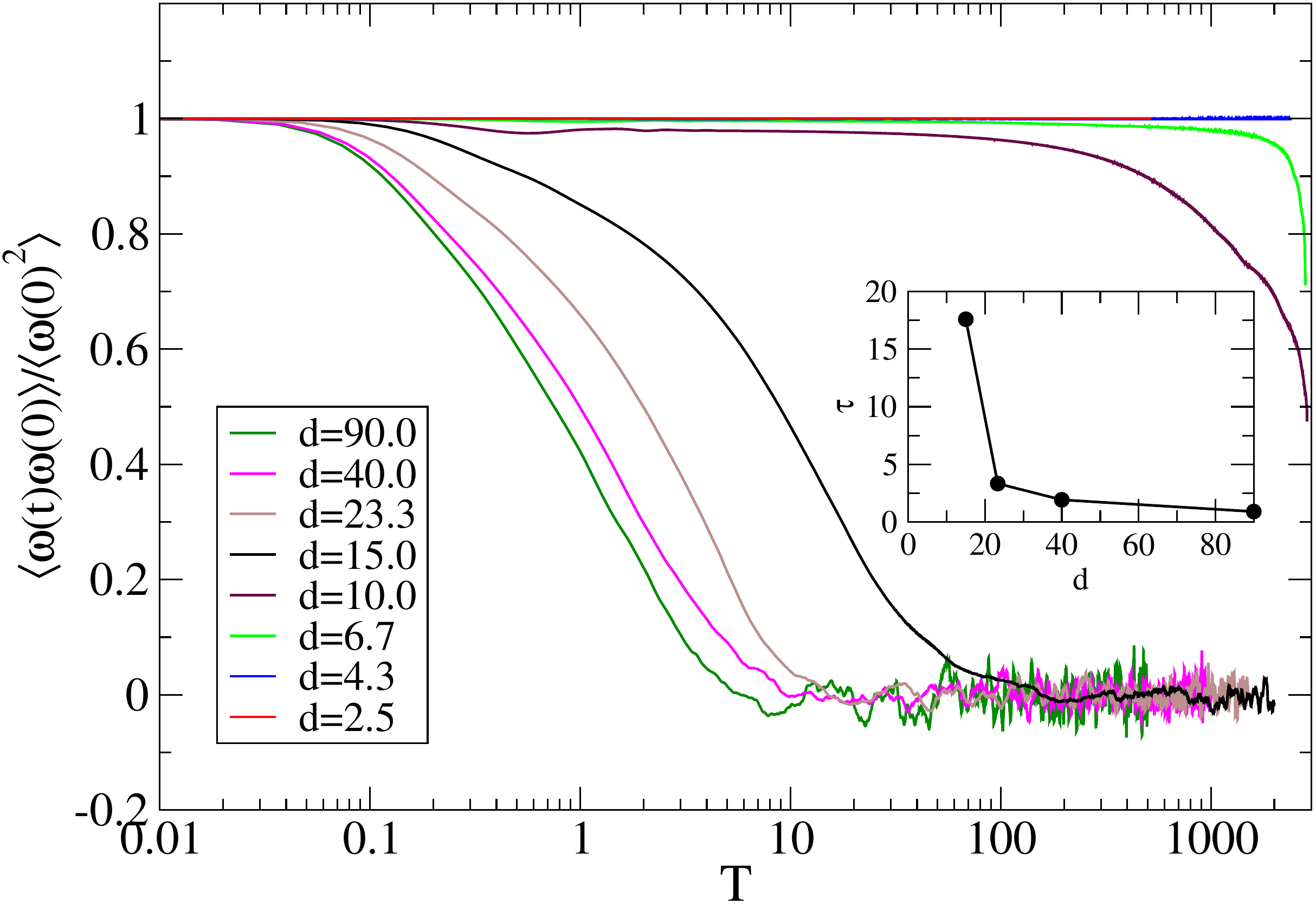}} \label{fig1c}
	\caption{{\bf Emergence of anti-ferromagnetic spin state in an array of rotating discs.} {\bf(a)} An array of counter-rotating discs in mesoscale turbulence generated by active matter. Rotors have a radius $R=50$ and are fixed on a square lattice with cell edge $100$. The director field is shown by solid green lines and is overlaid by streamlines (black lines with arrows). Discs are coloured according to their angular velocity as shown in the colourbar (red- anticlockwise, blue-clockwise). A $3\times3$ portion of an $8\times8$ lattice of rotors is shown (\mov{S1} and \mov{S2}). {\bf(b)} Cross correlations, $\left\langle \omega^i\left(t\right)\omega^j\left(t\right) \right\rangle$, between neighbouring rotors $(i,j)$ as a function of the minimum gap size $d$ between rotors for {\bf(i)} homeotropic anchoring {\bf(ii)} no anchoring and {\bf(iii)} planar anchoring of the director at the rotor surface. The root mean square (rms) angular velocity, $\omega_\textmd{rms}=\sqrt{ \left\langle \omega\left(0\right)^2 \right\rangle }$, is shown in the inset. {\bf(c)} Time correlations of rotor angular velocity $ \left\langle \omega\left(t\right)\omega\left(0\right) \right\rangle $, normalised by its mean value $\left\langle \omega\left(0\right)^2 \right\rangle $ for different lattice sizes as a function of scaled time $T=t/\left(2\pi/\omega_\textmd{rms}\right)$ for the case with homeotropic boundary conditions on the rotor surface. The decorrelation time $\tau$ is shown as a function of gap size in the inset for $d>10$.}
	\label{fig1}
\end{figure*}

We use numerical simulations to demonstrate that mesoscale turbulence can be exploited to generate work by considering a 2D square array of cylindrical rotors that are free to turn in response to mesoscale flows in the solution. The array of discs achieves a state of permanent rotation below a given inter-disc spacing (\fig{fig1}{(a)}). The spin-state of the rotors is fixed, with neighbouring discs rotating in opposite directions.  Hence, the disc array rectifies the active turbulence into a steady power source.

To determine how the antiferromagnetic spin-state of the rotors is established, we consider a solitary disc immersed in an active bath.  Protracted, but non-permanent, rotation can be achieved by synergistic hydrodynamic and elastic effects, which depend both on the properties of the active flow and on the orientational anchoring of the active fluid at the disc surface. We find an exponential dependence of the persistent rotation time on rotor size, which suggests that decorrelation of rotor angular velocity is controlled by an activated barrier-crossing process. In addition, by placing a single rotor within an immobile circular cavity, we explain how rotation stabilisation results from fluid confinement between neighbouring discs in the lattice configuration and demonstrate that the characteristic length of activity-induced vortices sets the optimal length scales for the microfluidic design.

\section{Results}

A square lattice of 64 rotors is deployed in simulations with periodic boundary conditions on domain boundaries and either homeotropic (perpendicular to the boundary), planar (parallel to the boundary) or no anchoring of the director (characterising the orientation field) at the rotor surface.  Drag on the rotors due to spontaneous active flows causes them to spin (\fig{fig1}{(a)}). Each disc preferentially rotates in the opposite direction to its nearest neighbours.  Although \fig{fig1}{(a)} shows only a subset of rotors, the alternating pattern repeats throughout the entire system (\mov{S1} and \mov{S2}). This strong anti-correlation between angular velocities occurs when the gap size between nearest-neighbours $d$ is small. However, if the spacing is increased, the spins of neighbouring rotors de-correlate and rotors spin randomly (\fig{fig1}{(b)}).

\begin{figure*}
	\centering
	\includegraphics[width=0.95\textwidth]{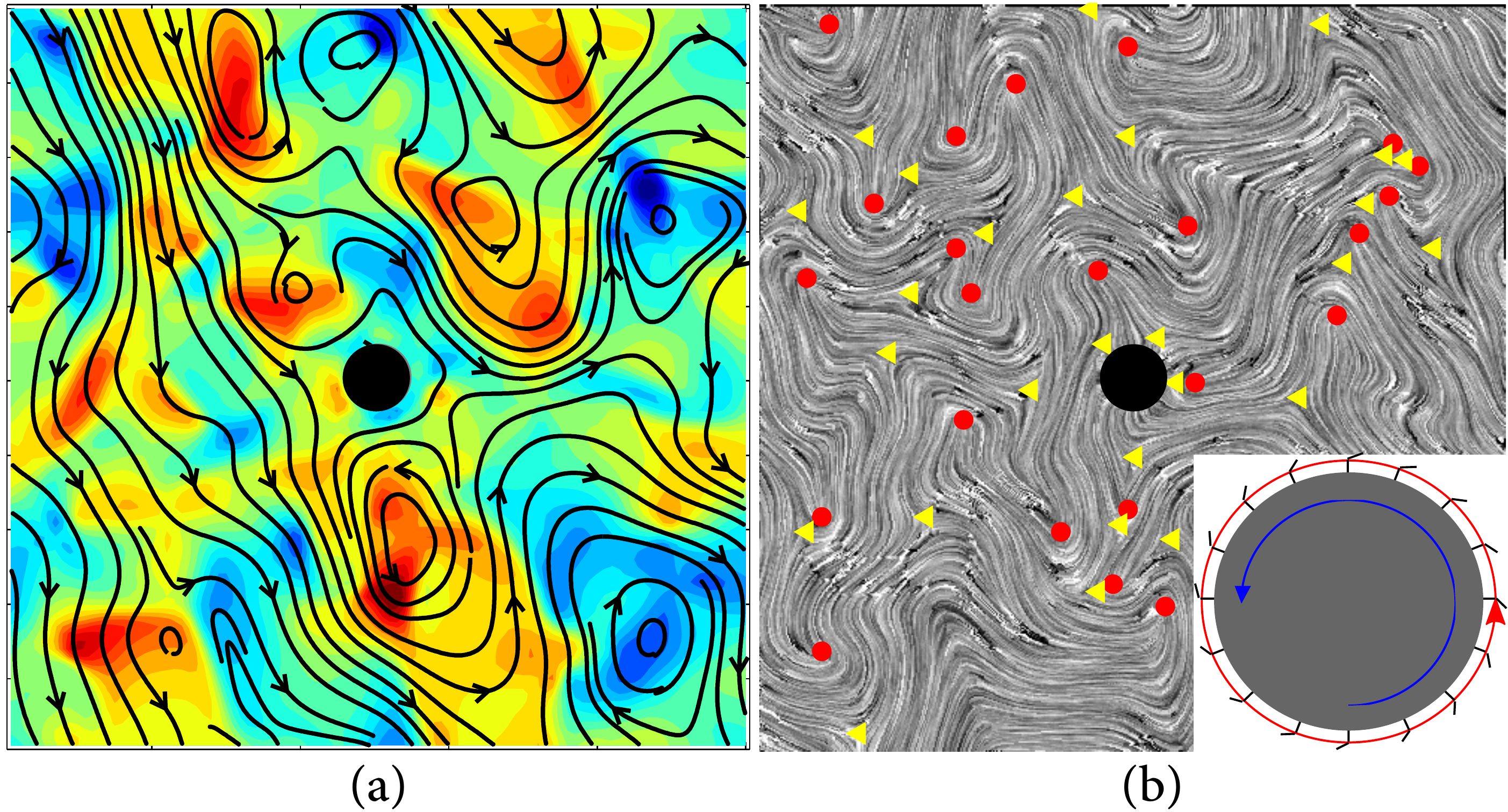}
	\caption{{\bf Rotation of a single rotor immersed in a turbulent flow of active matter with homeotropic anchoring.} {\bf(a)} Colourmaps show the vorticity contours and solid lines represent streamlines. Vorticity is normalised to its maximum value, blue and red representing clockwise and anti-clockwise rotation respectively. An animated version is available (\mov{S3}). {\bf(b)} Director field around the rotating disc. Red dots and yellow triangles denote $+\tfrac{1}{2}$ and $-\tfrac{1}{2}$ topological defects, respectively. \textit{Inset:} Schematic of the wall in the director field that encircles each rotor and the resulting active flow of fluid along the wall. }
	\label{fig2}
\end{figure*}

Moreover, the rotation of a single disc in the array is extremely enduring (\fig{fig1}{(c)} and \mov{S3}).  When the separation between rotors is small, it is perpetual on the time scale of our simulations ($10^7$ simulation time steps, which covers over 1000 rotations of the rotors) but, when the spacing increases, the angular velocity auto-correlation function of a single rotor, $\left\langle \omega\left(t\right) \omega\left(0\right) \right\rangle$,  goes to zero over time (\fig{fig1}{(c)}).  The auto-correlation function describes the propensity of the rotor to continue rotating in its current direction and is characterised by the time scale $\tau$. Even for the largest spacings, the rotors maintain a non-zero persistence time of rotation. 

Streamlines in \fig{fig1}{(a)} demonstrate that local flows move in the same direction as each rotor spins for small gap sizes.  The array of spinning rotors stabilises the active fluid flow, which would otherwise exhibit mesoscale turbulence.  Simulations that consider the extreme case of infinite friction (by disallowing disc rotation) show the same streamlines and director fields as the arrays of free rotors, though with reduced flow speeds. This indicates that the self-organised spin-state should be expected in experimental realisations of microfluidic systems of rotors that include finite rotational friction or have been designed to act as active turbines.

In order for a free rotor to change its direction of rotation in the small separation regime, it must overcome the local flow field, which is established not only through the boundary conditions of the disc in question but also entrenched by the flow field of neighbouring rotors. Although, the decorrelation time of the angular velocity increases with decreasing gap size, the rotation rate is non-monotonic with a sharp maximum (\fig{fig1}{(b)}; inset). The maximum is set by the competition between gap size and the active-nematic fields.  To better understand the mechanisms leading to steady anti-correlated rotation and the impact of rotor spacing, we systematically investigate the dynamics of a solitary rotor and a single rotor within a concentric cavity. These single-rotor results demonstrate how the permanent spin-state arises in the array and explain how the optimal rotation rate relates to the intrinsic characteristic vorticity length scale of the active turbulence.

A solitary disc in an active fluid begins to rotate once activity-driven flow fields are established (\fig{fig2}{\null}).  The surrounding active fluid exhibits self-sustained mesoscale turbulence (\fig{fig2}{(a)}) even at zero Reynolds number~\cite{Jorn2013}.   The chaotic vorticity field is driven by a ‘reverse energy cascade’ of energy produced at microscopic length scales escalating to larger-scale wavelengths~\cite{Giomi2014turb,Bratanov2015}. In active-nematic fluids, the continual fluid motion is coupled to the formation of lines of strong bend distortions in the orientation field, referred to as a wall~\cite{ourepl2014,Shelly2014}, and their unzipping through the creation and annihilation of topological defect pairs~\cite{Giomi2014,ourpta2014}. The resulting stochastic velocity and vorticity fields have been noted to exhibit many features of turbulent flow~\cite{Julia2012,Jorn2013,creppy2015} and can be characterised by the vorticity-vorticity correlation length scale.

\begin{figure*}
	\centering
	\subfigure[]{\includegraphics[width=0.45\textwidth]{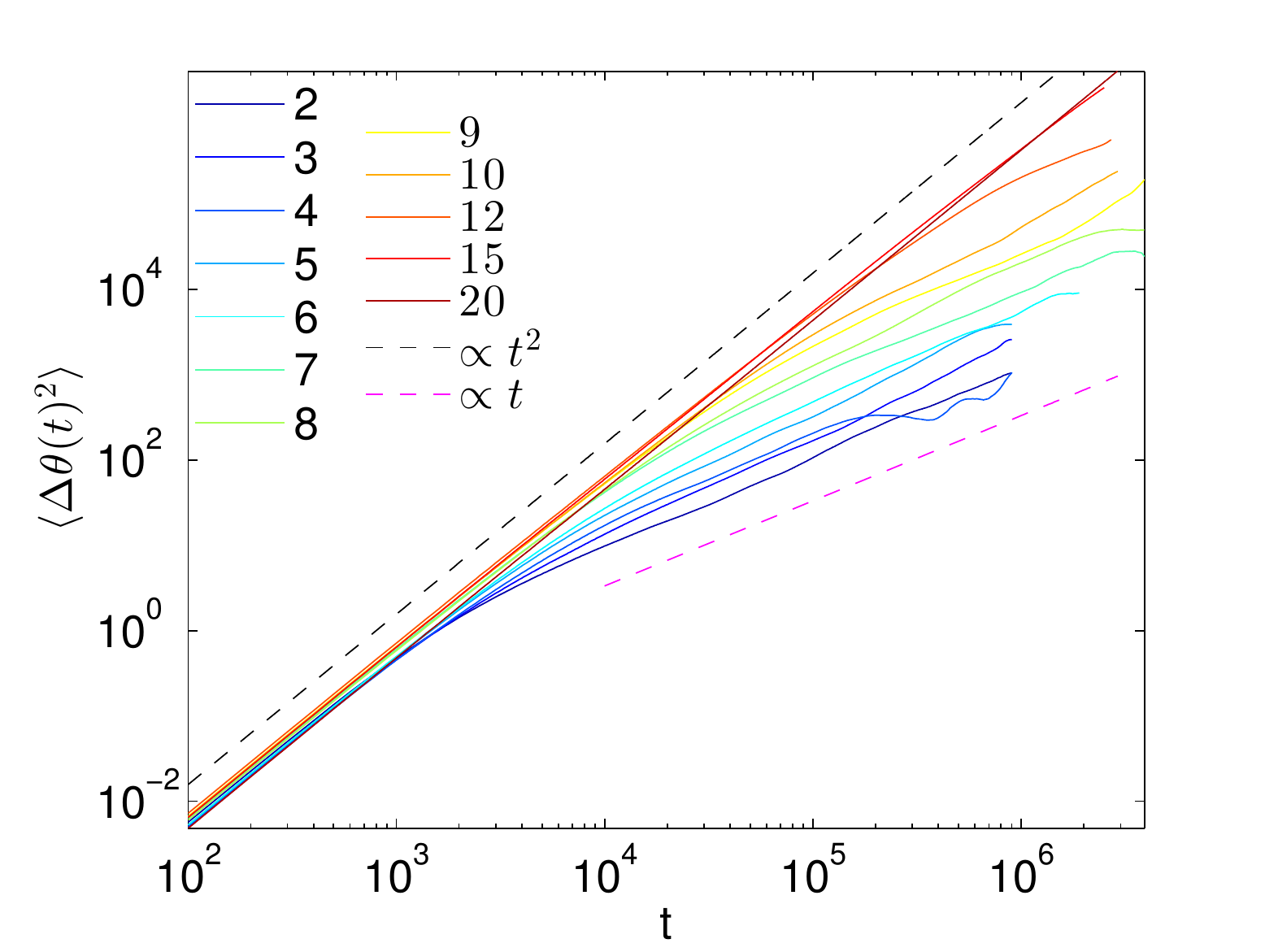}} \label{fig3a}
	~
	\subfigure[]{\includegraphics[width=0.45\textwidth]{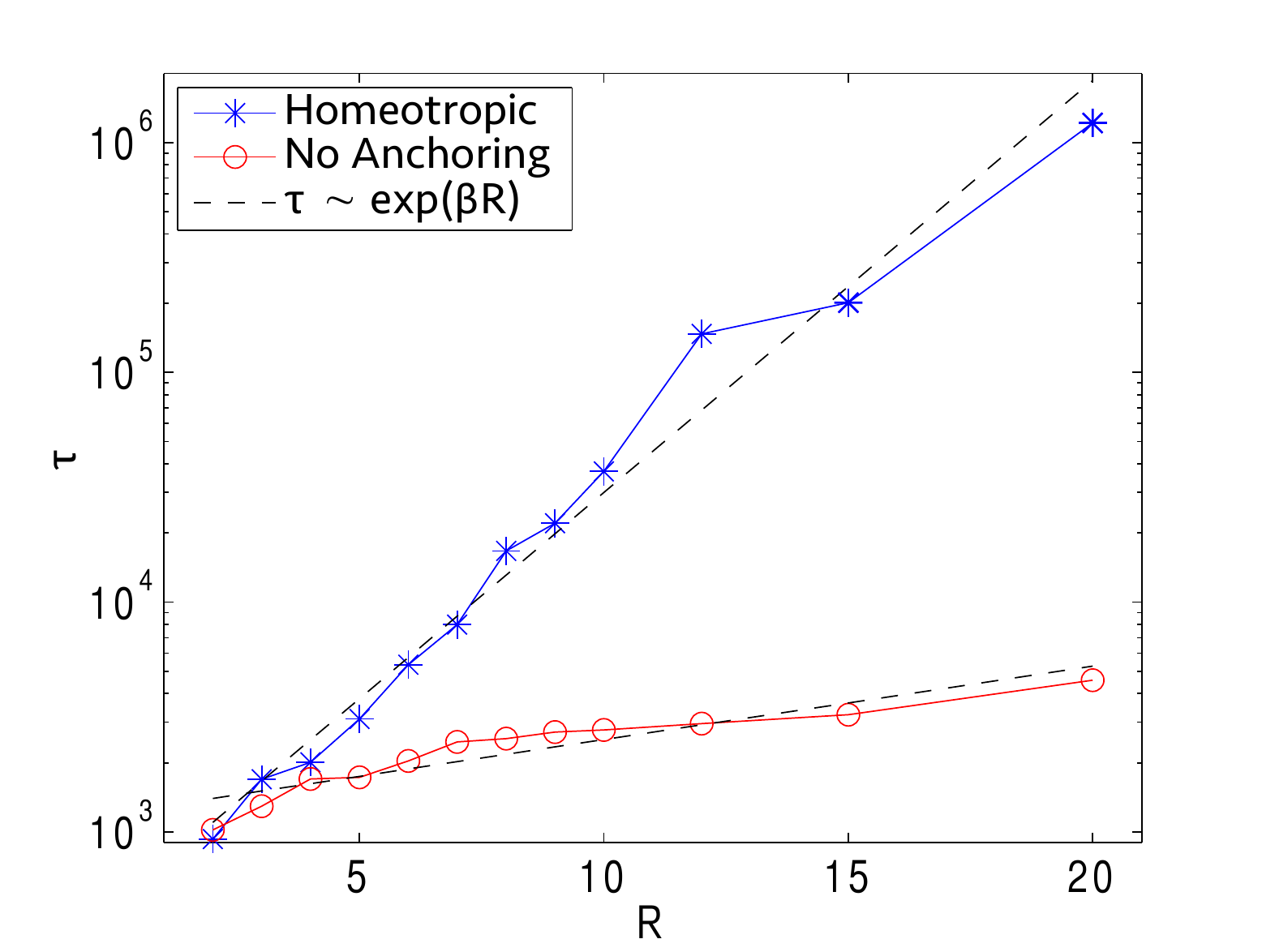}} \label{fig3b}
	\caption{{\bf Rotational dynamics of a single rotor in mesoscale turbulence.} {\bf(a)} Mean-squared angular displacement of rotors versus time for different rotor radii $R$. Director boundary conditions on the rotor surface are homeotropic. {\bf(b)} Persistence time as a function of $R$ comparing homeotropic and no anchoring director boundary conditions on the rotor surface.}
	\label{fig3}
\end{figure*}

The rotational dynamics are controlled by both the nature of the mesoscale turbulence and the rotor characteristics. Spontaneous symmetry breaking randomly establishes clockwise or counter clockwise spin and thereafter the rotor sporadically switches direction. Hence, as shown in \fig{fig3}{(b)}, at short times the rotation of the disc in response to the flow is propulsive
\begin{align}
 \left\langle \Delta\theta\left(t\right)^2 \right\rangle &= \omega_\textmd{rms}^2 t^2, \qquad\qquad t\ll\tau. 
\end{align}
The short-time behaviour is referred to as propulsive rather than ballistic because rotor inertia is viscously damped and rotation is due to the nematohydrodynamic propulsion from the active flow field. At longer times there is a crossover to a diffusive regime
\begin{align}
 \left\langle \Delta\theta\left(t\right)^2 \right\rangle &= 2D_\textmd{r} t, \qquad\qquad t\gg\tau, 
\end{align}
where $\omega_\textmd{rms}$ is the root-mean-square angular velocity and $D_\textmd{r}$ is an effective rotational diffusion constant. The diffusive regime arises as active vortices impinge on the rotor and so is controlled by the characteristic decay time of the vorticity-vorticity correlation function $D_\textmd{r}\sim\tau^{-1}$.  The decay time $\tau$ is substantially prolonged by increasing the size of the rotor as is apparent both from \fig{fig3}{(b)} and from the decay of the angular velocity auto-correlation function.

The rotation is driven by two synergistic mechanisms:
\begin{enumerate}
 \item entrainment of an active turbulence vortex around the rotor
 \item flow driven by the existence of a wall at the surface of the rotor (\fig{fig2}{(b)}; inset).
\end{enumerate}
The second mechanism arises due to the orientational boundary condition on the surface of a disc. For homeotropic anchoring, bend deformations in the director field appear naturally at the surface~\cite{Joanny2005,Davide2007}. The bend deformations result in the formation of a wall around the perimeter of each rotor~\cite{ourepl2014,Shelly2014} (\fig{fig2}{(b)}; inset).  Simulations of colloids pulled through active nematic fluids have previously shown that orientational distortion can lead to local flows causing non-Stokesian dynamics and even negative drag~\cite{Cates12}. Likewise, in this system the rotor-encircling wall establishes a unidirectional flow~\cite{ourepl2014} that generates localised vorticity (\mov{S3}). Supporting this argument, if, instead of homeotropic anchoring, no anchoring boundary conditions are implemented, the persistence rotation time of the rotor is significantly diminished since no wall is formed at the disc surface (\fig{fig3}{(c)} and \mov{S4}). The wall-generated flows spontaneously establish a direction of rotation and any co-rotational vortices enhance the rotational velocity when the characteristic vortex size of the mesoscale turbulence is commensurate to the rotor size. Thus, the unidirectional flow generated within the wall and the swirling flows within active vortices collude to give a persistent rotation of the disc.

\begin{figure}
	\centering
	\subfigure[]{\includegraphics[width=0.225\textwidth,trim={0.75cm 6cm 8.25cm 0},clip]{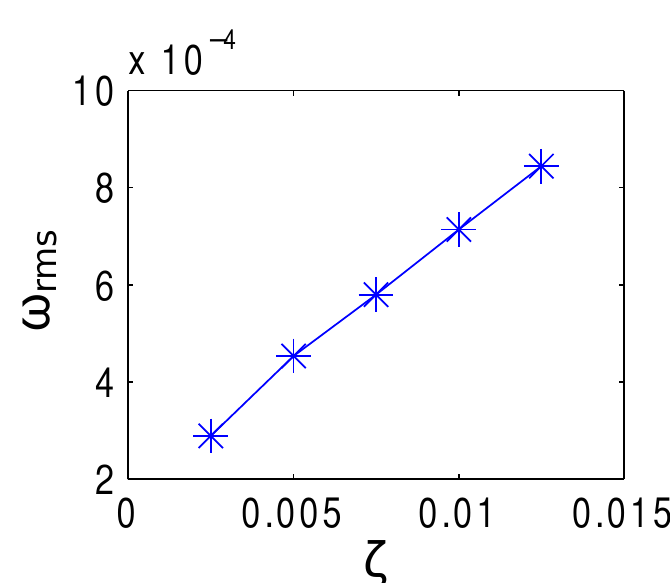}} \label{fig4a}
	~
	\subfigure[]{\includegraphics[width=0.225\textwidth,trim={0.75cm 6cm 8.25cm 0},clip]{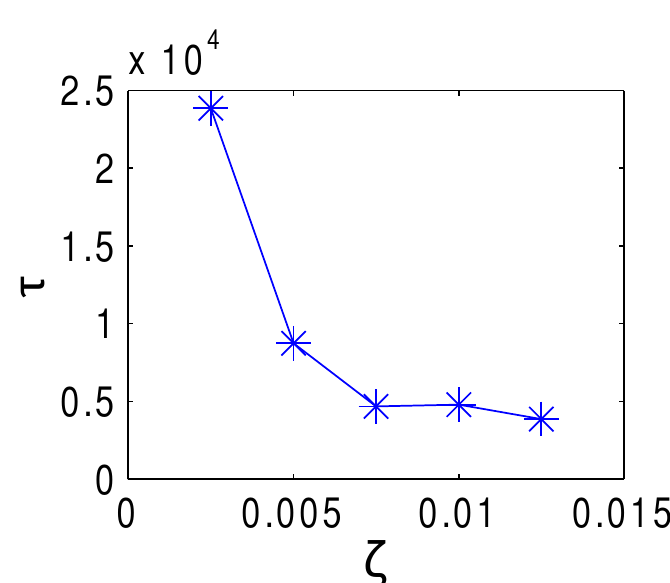}} \label{fig4b}
	\caption{{\bf Characterisation of a single rotor in mesoscale turbulence.} {\bf(a)} Increasing activity results in an enhanced angular velocity. {\bf(b)} Increasing activity shortens the rotational persistence time. }
	\label{fig4}
\end{figure}

Extracting the decay times from the autocorrelation functions and plotting them against $R$ shows that $\tau$ grows exponentially with the radius of the rotor (\fig{fig3}{(c)}). The exponential dependence of rotational persistence time on rotor radius suggests an activated barrier crossing, in which flipping a rotor’s spin is a stochastic process over an energy barrier that is linearly dependent on rotor size. The barrier that must be overcome by local fluctuations in the vorticity field in order to de-correlate the propulsive rotation is the reorientation of the wall around each rotor, with a total energy proportional to the perimeter $\sim R$. In this way, spin flipping is analogous to a non-thermal activated Kramers process with  $\tau \sim e^{\beta R}$ where $\beta$ is a non-thermal constant. Once the wall flips direction, it drives a reversal of the unidirectional flow in the immediate vicinity of the surface. In this way, the intrinsic active-elastic properties of the nematic establish a barrier to active turbulence flipping the rotor spin. As activity is increased, the rotation rate increases linearly (\fig{fig4}{(a)}). However, the increasingly energetic fluctuations in the mesoscale turbulence are more able to propel the system over the energy barrier between different directions. Hence, the correlation time decreases with activity (\fig{fig4}{(b)}).

\begin{figure}
	\centering
	\subfigure[]{\includegraphics[width=0.45\textwidth,trim={0 5.25cm 0 0},clip]{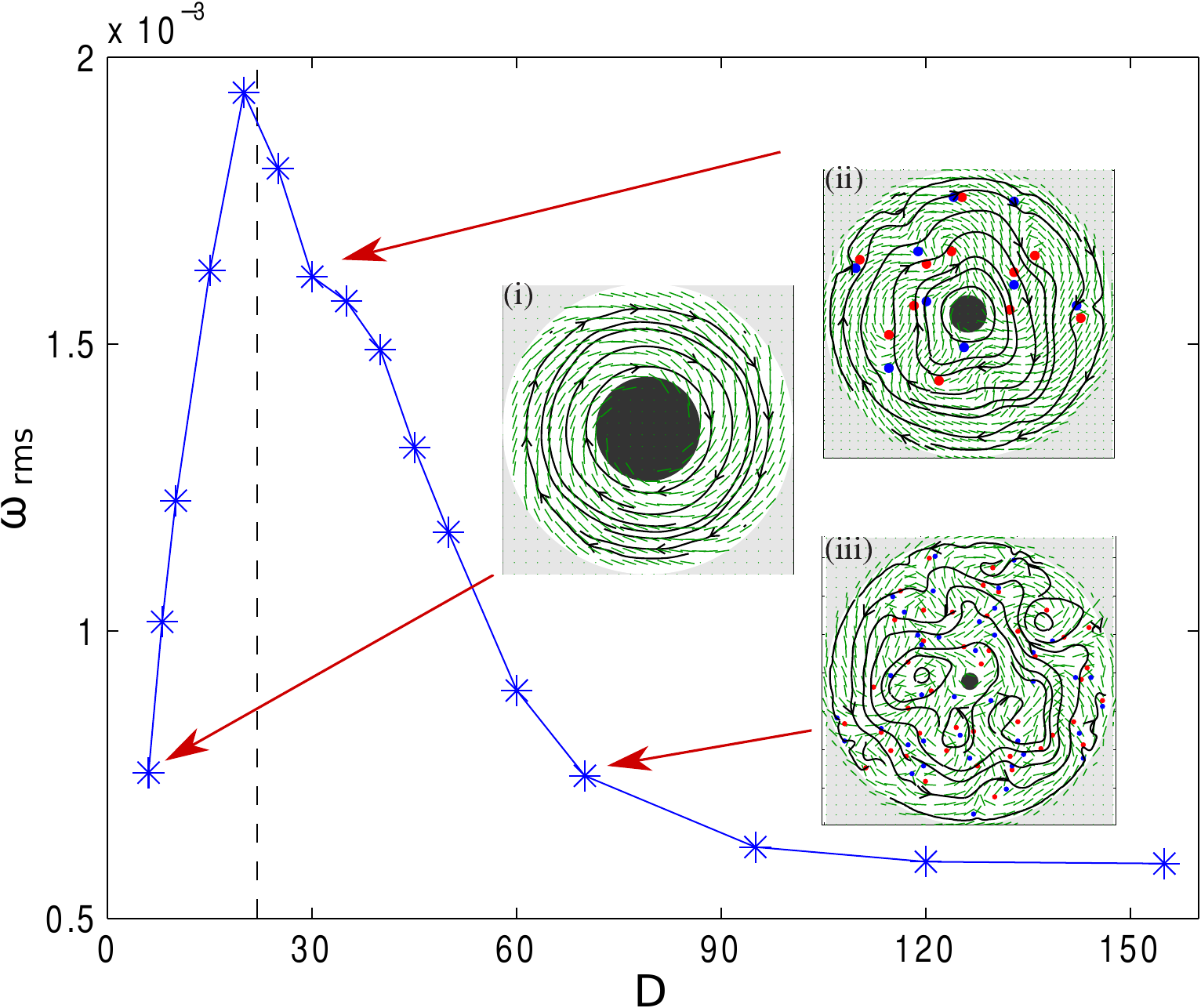}} \label{fig5a}
	\\
	\subfigure[]{\includegraphics[width=0.45\textwidth]{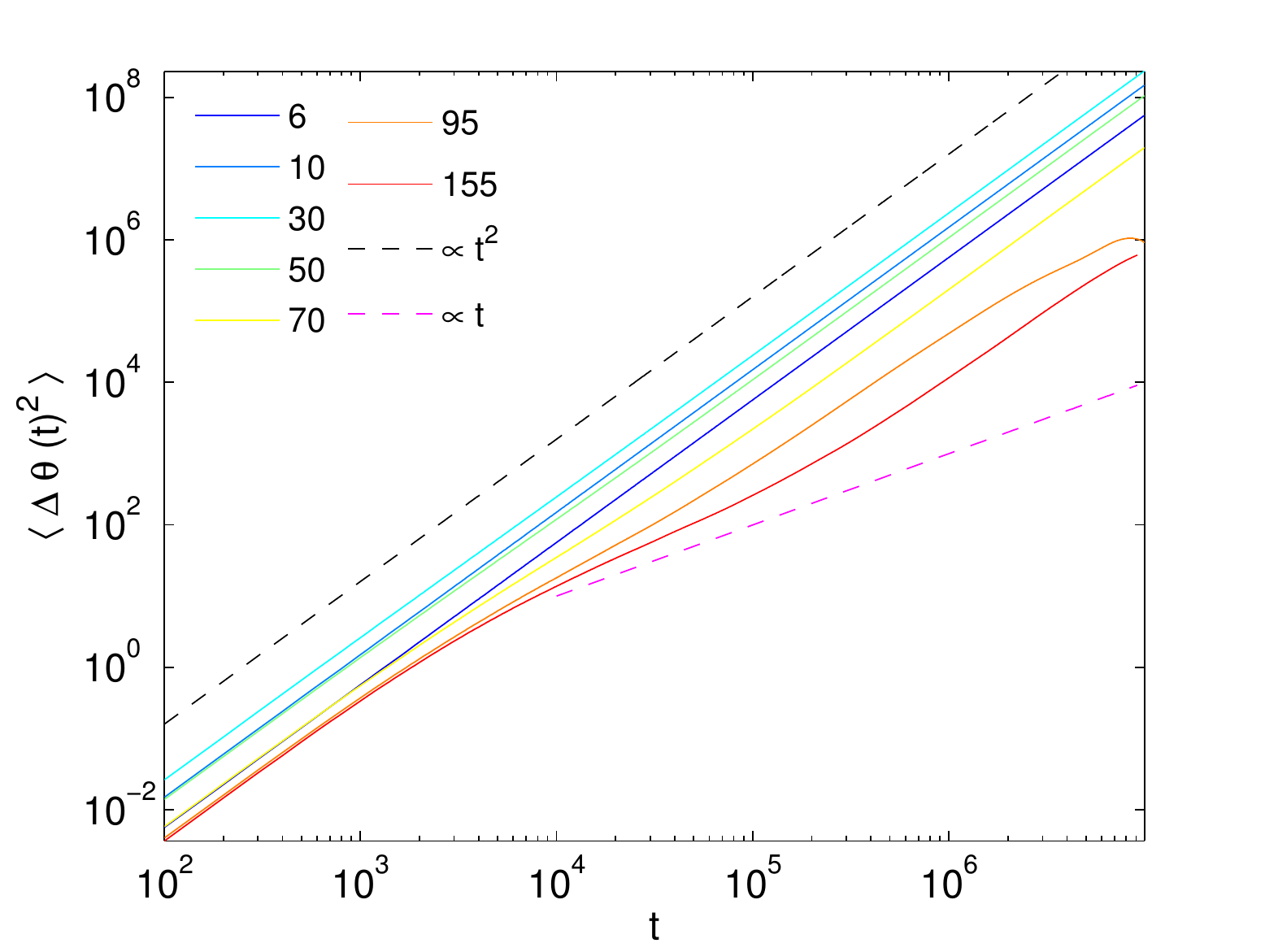}} \label{fig5b}
	\caption{{\bf Confinement-induced rotation of a disc immersed in a bath of active matter showing the transition from unidirectional to turbulent flow.} {\bf(a)} The main panel shows the variation of the angular velocity of the rotor with increasing gap size. The vertical dashed line corresponds to the characteristic length scale of bulk mesoscale turbulence measured from the vorticity-vorticity correlation function in bulk. The insets show the transition from unidirectional flow to active turbulence with increasing gap size $D$. {\bf(i)} $D=8$, {\bf(ii)} $D=30$, {\bf(iii)} $D=70$. Shown here is the director field (dashed green lines), superposed by streamlines (solid black lines with arrows) and topological defects (red and blue dots denoting $+\tfrac{1}{2}$,$-\tfrac{1}{2}$  defects, respectively). {\bf(b)} Mean square angular displacement of the rotor as a function of simulation time, where the transition from propulsive to diffusive behaviour is seen when gap size $>70$.}
	\label{fig5}
\end{figure}

A solitary symmetric rotor ensnares a vortex and establishes a barrier to de-correlation, leading to large persistence times but not as persistent a spin-state as observed for arrays of discs. We now ask why the array of rotors can maintain such a stable spin-state. We consider a single disc within a fixed circular cavity (\fig{fig5}{\null}). Several authors have shown that confinement can stablise active flows and leads to spontaneous circulations~\cite{Joanny2005,Goldstein2012,Miha2013}. Even in the absence of a prescribed orientation at the surfaces, the active fluid self-organises within the confinement of the annulus to induce a unidirectional flow, which leads to an enduring rotation of the inner cylindrical rotor (\fig{fig5}{(a)}). As the gap size  $D$ between the rotor and confinement boundary increases from small values, the angular velocity of the rotor increases (\fig{fig5}{\null}). The rotation rate continues to increase with gap size until defects and distortions appear within the orientation field (\fig{fig5}{(b)}), at which point the rotation rate crests and begins to decrease. The maximum corresponds to the gap size becoming equal to the characteristic length scale of vorticity in bulk active turbulence, measured from the decay of the vorticity autocorrelation function (\fig{fig5}{\null} and \mov{S5}).

The maximum in the rotation rate is also marked by the appearance of topological defects in the director field. These leave the streamlines relatively unperturbed, with the defects advected in a laminar manner (\fig{fig5}{(b)}). However, for still larger gaps, obvious vortices appear in the flow field (\fig{fig5}{(c)}) and the rotor rotation rate slowly decreases with increasing $D$ (\fig{fig5}{\null}).

The rotational persistence time of the rotor is finite at large gap sizes with multiple vortices but is substantially enhanced when only a single stabilised vortex fits within the confinement. On the other hand, dissipation is increased when the gap size is smaller than the natural size of the vortices. From this, we understand the persistent counter-rotating spin-state of a disc array (\fig{fig1}{(a)}): Each rotor establishes a wall-induced active shear flow, while neighbouring rotors do the same, simultaneously creating a confinement that stabilises the local flow. This process is most efficient if the spacing between rotors is comparable to the characteristic vorticity length scale of the mesoscale turbulence (\fig{fig5}{\null} and \mov{S5}). In typical bacterial suspensions in 2D geometries, the characteristic length scale is $\sim 40\mu$m~\cite{Jorn2013}.

\section{Discussion}

We have demonstrated the persistent, correlated rotation of an array of symmetric rotors immersed in a dense suspension of active matter, such as bacteria suspensions~\cite{Julia2012}, filament-motor protein solutions~\cite{guillamat2016} or cellular monolayers~\cite{ourSM2015}. The results show that harnessing useful work from the disordered and chaotic flows of active materials does not require specifically designed asymmetric microscopic gears or external fields.

The persistent spin-state of the rotor array is antiferromagnetic, similar to the recently observed spin-states of bacteria suspensions confined within interconnected cavities~\cite{wioland15}. However, the physical mechanisms leading to these particular spin-states are markedly different. While the rotation of B. subtilis in interconnected circular cavities is dictated by a counter-rotating single cell layer that is controlled by the size of the intersection compared to the cell size~\cite{wioland15}, the antiferromagnetic spin-state of rotors studied here arises through a combination of an activated barrier-crossing process and mooring active nematic vortices to each rotor. The array of spinning rotors stabilises the active fluid flow, which would otherwise be in a state of mesoscale turbulence.

The way the rotor lattice constrains active excitations of the vorticity field, which in turn causes nonzero torques to arise on the rotors themselves, bears a strong conceptual similarity to Casimir forces (as defined in the general sense and not just for the electromagnetic field fluctuations)~\cite{Ramin1999}. The notion of using this concept to design micro-machines has been explored in the context of the traditional Casimir forces, revealing interesting possibilities afforded due to lack of physical contact~\cite{Ramin2007,Ramin2008}. A key aspect of the Casimir effect is that it depends primarily on geometric characteristics of the system, and as such offers a simple route to tuning the behaviour of the micro-machine. We note that Casimir forces have been postulated to exist in active systems~\cite{Ray2014,parrarojas2014}.

By investigating the dynamics of solitary rotors, we determine how different length scales, namely the rotor size, lattice spacing and the gap between the rotors, affect the behaviour of the array of rotors. We find that in larger gaps de-correlating vortices can be excited between rotors, while in smaller spaces counter rotating vortices around neighbouring discs viscously dissipate energy. Additionally, the spin-state of each rotor is locked by an energy barrier due to the thin active-nematic wall that forms on its surface and larger rotors have longer perimeters and higher barriers that further stabilise the spin-state. Thus, the choice of configuration and size of rotors is specific to a microfluidic design, and these choices can be optimised using vorticity correlation length as the characteristic length scale of the fluid.

Our findings demarcate a promising direction towards harnessing power from active matter. The ability to exploit the continuous injection of energy from constituent elements of active matter to obtain an ordered lattice of counter-rotating rotors has implications in many applications, from providing driving power for MEMS devices to acting as novel microfluidic mixers. 

\section{References}
\bibliography{refe}
\bibliographystyle{unsrt}

\section{Materials and Methods}

We use the equations of active nematohydrodynamics based on the theory of liquid crystals, which have proven successful in describing spatio-temporal dynamics of active matter systems, including bacterial suspensions~\cite{Volfson2008}, microtuble bundles~\cite{Dogic2012,Giomi2013,ourprl2013} and cellular monolayers~\cite{ourSM2015,Julicher2008}. The orientational order of microscopic active particles is represented by the nematic tensor $\tens{Q}=\tfrac{3q}{2}\left( \vec{n}\vec{n}-\tens{I}/3 \right)$, with $q$ the magnitude of the orientational order,  $\vec{n}$ the director, and  $\tens{I}$ the identity tensor, which evolves due to a co-rotation term and relaxation through rotational diffusivity $\Gamma$ of the molecular field, which accounts for both the Landau-de Gennes bulk free energy and the nematic distortion free energy~\cite{DeGennesBook}, assuming a single elastic constant $K$. The total density and the velocity field $\vec{u}$ of the active matter obey the incompressible Navier-Stokes equations, with a stress tensor $\tens{\Pi}$ that must account for contributions from the viscosity $\eta$, the elastic stress (which includes the pressure $P$~\cite{Berisbook}), and the active contribution to the stress $\tens{\Pi}^\textmd{act}=-\zeta \tens{Q}$~\cite{Sriram2002}.  Any gradient in $\tens{Q}$ generates a flow field through the active stress, with strength determined by the activity coefficient, $\zeta$.

The equations of active nematohydrodynamics are solved using a hybrid lattice Boltzmann technique~\cite{ourpta2014} that does not include thermal fluctuations. The momentum equation is solved using the lattice Boltzmann method to resolve the hydrodynamics.  Method of lines~\cite{ourpta2014} is implemented to determine the order parameter, in which a finite-difference approach is used for spatial discretisation, and the temporal evolution is obtained through an Euler integration scheme. For details, see~\cite{ourpta2014,Davide2007,Suzanne2011}.

As usual in lattice Boltzmann schemes, discrete space and time steps are chosen as unity and all quantities can be converted to physical units in a material dependent manner~\cite{Giomi2013,Cates2008,Henrich2010}. The parameters used in the simulations are $\Gamma=0.34$, $K=0.01$, $\zeta=0.01$ and $\eta=2/3$, in lattice units. These parameters result in active turbulence in the bulk. Unless otherwise stated, simulations were performed in a 2D domain of size $200\times200$. Microrotors are modeled as discs discretized on the simulation lattice. Each rotor is fixed in space but allowed to spin freely with a moment of inertia $I=10^3$ for all radii. For the parameters used here, viscous damping dominates over inertia in approximately $\sim I/\eta R^2 \sim 10^0$ time steps. No-slip boundary conditions allow flows to apply torques to the rotors. 

\section{Supplementary Information}

\begin{description}
 \item{\mov{S1}:} {\bf Anti-ferromagnetic spin state in an array of rotating discs.} Rotors have a radius $R=5$ and are fixed on a square $8\times8$ lattice with cell edge $100$. The magenta arrows show the instantaneous orientation of each disc. Vorticity of the active nematic is shown in the colourbar, which is normalised by its maximum value, with blue and red representing clockwise and anti-clockwise rotation respectively. 

 \item{\mov{S2}:} {\bf Anti-ferromagnetic spin state in an array of rotating discs.} A $3\times3$ portion of the micromachine shown in {\bf \mov{S1}}. 

 \item{\mov{S3}:} {\bf Rotation of a single rotor immersed in active turbulence.} The director field obeys homeotropic anchoring at the rotor surface. Colourmap shows the instantaneous vorticity contours and solid lines represent streamlines. Vorticity is normalised to its maximum value, blue and red representing clockwise and anti-clockwise rotation respectively. Localised vorticity causes the persistent rotation of the single disc. 

 \item{\mov{S4}:} {\bf Rotation of a single rotor immersed in active turbulence.} Same as {\bf \mov{S3}} but with no anchoring boundary conditions at the surface of the rotor. 

 \item{\mov{S5}:} {\bf Confinement-induced rotation of a disc immersed in an active nematic bath.} Colourmap shows the instantaneous vorticity contours and solid lines represent streamlines. The gap size $D=30$ is comparable to the characteristic length scale of vorticity in bulk active turbulence, measured from the decay of the vorticity autocorrelation function, such that only a single active vortex can reside in the gap. 
\end{description}

\section{Acknowledgments}

\paragraph{General:} The authors are grateful to Mitya Pushkin for insightful discussions.

\paragraph{Funding:} This work was supported through funding from the ERC Advanced Grant 291234  MiCE and we acknowledge EMBO funding to TNS (ALTF181-2013). 

%
%

\end{document}